\documentclass{aa}  
\usepackage{graphicx}
\usepackage{txfonts}
\usepackage{natbib}
\usepackage{stfloats}
\makeatother

\def\starlight{{\sc Starlight}}
\def\fado{{\sc Fado}}
\def\dyn{{\sc Dynamite}}
\def\ppxf{{\sc pPXF}}

\def\D4000{$D_{4000}$}

\def\RY{${\cal RY}$}

\def\mbstar{${\cal M}_{\star,\textrm{B}}$}
\def\mstotal{${\cal M}_{\star,\textrm{T}}$}

\def\mtmass{$\langle t_{\star,\textrm{B}} \rangle_{{\cal M}}$}
\def\mZmass{$\langle Z_{\star,\textrm{B}} \rangle_{{\cal M}}$}

\usepackage{hyperref}
\def\mdtmass{$\langle t_{\star,\textrm{D}} \rangle_{{\cal M}}$}

\begin{document} 

% =============================================================================================
\title{Tracing the dynamical and structural complexity of spiral galaxy centres}
% =============================================================================================
   \author{
          Iris Breda \inst{\ref{UniVie}}
          \and
          Glenn van de Ven \inst{\ref{UniVie}}
          \and
          Sabine Thater \inst{\ref{UniVie}}
          \and
          J. Falc\'on-Barroso \inst{\ref{IAC}, \ref{IAC0}}
          \and
          Prashin Jethwa \inst{\ref{UniVie}}
          \and
          Masato Onodera \inst{\ref{Tokyo}, \ref{Subaru}}
          \and
          Joop Schaye \inst{\ref{Leiden}}
          \and
          Jarle Brinchmann \inst{\ref{UPorto}}
          \and
          Bodo Ziegler \inst{\ref{UniVie}}
          \and
          Federica Mauro \inst{\ref{UniVie}}}
          
\institute{Dep. of Astrophysics, University of Vienna, Türkenschanzstraße 17, 1180, Vienna, Austria \label{UniVie}
         \and
Instituto de Astrof\'isica de Canarias, Calle V\'ia L\'actea s/n, E-38205, La Laguna, Tenerife, Spain \label{IAC}
                \and
Dep. de Astrof\'isica, Universidad de La Laguna, Av. del Astrof\'isico Francisco S\'anchez s/n, E-38206, La Laguna, Tenerife, Spain \label{IAC0}
                 \and 
Centre for Extragalactic Astronomy, Department of Physics, Durham University, South Road, Durham DH1 3LE, UK \label{CEA}
         \and
Graduate Institute for Advanced Studies, SOKENDAI, 2-21-1 Osawa, Mitaka, Tokyo 181-8588, Japan \label{Tokyo}
         \and
Subaru Telescope, National Astronomical Observatory of Japan, 650 N Aohoku Pl, Hilo, HI96720, Japan \label{Subaru}
         \and
Dep. de F\'isica e Astronomia, Faculdade de Ci\^encias, Rua do Campo Alegre, 4169-007 Porto, Portugal \label{UPorto}
                 \and
Leiden Observatory, Leiden University, PO Box 9513, 2300 RA Leiden, the Netherlands \label{Leiden}  
\\
\email{iris.breda@univie.ac.at}
}

\date{Received ???; accepted ???}

\abstract
%Context
{The formation and evolution of late-type galaxies have traditionally been described in terms of two pathways, one producing pressure-supported classical bulges and the other producing rotationally supported pseudo-bulges. Early studies supporting this view were primarily based on photometric decompositions that assumed an exponential disk, extrapolating it inwards. However, recent high-resolution observations have revealed a far more complex landscape in disk galaxy centres.}
% Aims
{We investigated the morphology of central stellar components in a subset of intermediate-to-massive spiral galaxies at unprecedented detail, focusing on disentangling the contributions of their cold, warm, and hot orbital components. Our goal is to critically reassess the standard approach of extrapolating a disk's exponential profile into the galaxy centre.}
% Methods
{To this end, we developed the pipeline \textsc{GLANCE} (Galactic archaeoLogy via chronochemicAl \& dyNamiCal modElling), a dedicated tool for photometric, chronochemical, and dynamical galaxy analysis. We applied \textsc{GLANCE} to eight high-resolution MUSE galaxies, deriving stellar population properties, and decomposing their orbits into cold, warm, hot, and counter-rotating components.}
%Results
{We uncovered a remarkable structural diversity in the dynamically cold central component of our sample galaxies: While one galaxy displays a tentatively exponential profile throughout its full extent, the majority exhibit either a pronounced central drop resembling a doughnut-shaped structure or a compact inner disk that is significantly steeper than the outer parent disk. Surprisingly, most galaxies that host nuclear disks could be classified as classical bulges -- being dynamically hot, old, and red and having a high bulge-to-total ratio -- in contrast to the majority of galaxies that exhibit a central deficit in the cold component. On the other hand, the luminosity contribution of cold plus warm orbits beyond the bulge generally remains below the total, indicating that the parent disk contains a non-negligible fraction of hot or counter-rotating orbits, displaying radial profiles with varying S\'ersic indexes consistently larger than unity.}
%Conclusions
{Our analysis indicates that the centres of disk galaxies are often more complex than what is implied by a simple inwards continuation of the outer exponential disk profile. These results highlight the composite nature of central structures in disk galaxies and the need for decomposition techniques that do not indiscriminately rely on extrapolating the outer disk into the innermost regions. Development of such methods will require detailed studies of a statistically representative sample with high-quality integral field spectroscopy data spanning a broad mass range, ideally complemented with high-resolution simulations such as Illustris TNG50.}

\maketitle
%________________________________________________________________

%\parskip = \baselineskip

\section{Introduction \label{intro}}

To this day, the mechanisms governing the formation and subsequent evolution of late-type galaxies (LTGs) remain a controversial topic within the realm of extragalactic astronomy. Originally, galactic disks were thought to form early on via violent quasi-monolithic gas collapse \citep{Lar74,Bur87} or galaxy mergers \citep[e.g.][]{BarHer96,SprHer05,BouJogCom05} and/or to gradually assemble around pre-existing monolithically collapsed bulges \citep[e.g.][]{Kau93,Zoc06}. Later on, improved observational data were used  to identify galactic disks that do not possess a prominent central stellar excess, and a dichotomy on the classification of spirals was introduced, being now common practice to empirically subdivide spirals into two categories -- classical bulge (CB) versus pseudo-bulge (PB; see \citealt{KorKen04} for a review) -- primarily based on their bulge's S\'ersic index. Indeed, a large fraction of local LTGs host PBs \citep[e.g.][]{Gad09,FisDro11,FerLor14}, and their spectro-photometric and kinematical characteristics substantially differ from CBs. 
As these early works established a formation dichotomy in which CBs are interpreted as relics of violent early assembly while PBs are linked to long-term secular evolution, it was already recognised that real systems may span a continuum and that blue or disky features do not uniquely trace secular formation, as merger-built bulges may undergo subsequent rejuvenation \citep[e.g.][]{KorKen04,Gad09,ThoDav06,CoeGad11}. Nevertheless, this empirical dichotomy remains widely adopted as a working framework in recent observational studies \citep[e.g.][]{Wan19,Gao20,Hu24a,Hu24b}.

Segregation into one of the two categories implies a bimodal formation pathway for spirals: While the genesis of LTGs hosting CBs envisages a two-phase scenario by early-on formation of a kinematically hot bulge through violent quasi-monolithic collapse followed by gradual disk assembly, PBs are thought to form out of their parent disks through gentle gas inflow and subsequent in situ star formation (SF). However, a recent study based on photometric decomposition of 3106 galaxies, including all Hubble types \citep{Qui23}, reports a continuous transition -- from small, faint disk-like central concentrations to prominent spheroids -- rather than a clear dichotomy between CBs and PBs. In addition, current studies predicated on spatially resolved population spectral synthesis of a representative sample of the local population of spiral galaxies have uncovered tight and continuous correlations between all main properties of the bulge (namely, stellar mass, stellar density, bulge-to-total, mean age and metallicity, and formation timescale) and the total galaxy mass \citep{BP18,Bre20a,BP23}. 
Furthermore, this analysis has revealed that the main physical properties of bulges and their parent disks are closely related, strongly suggesting concurrent formation and co-evolution of these two stellar structures. 

This result is in sharp contrast with conclusions drawn from studies based on photometric decomposition, which frequently favour independent bulge-disk co-evolution scenarios \citep[e.g.][]{Mar16,Mar18},\footnote{There are, however, seemingly contradictory results, frequently resulting in conflicting interpretations \citep[see][for a review]{Bre20b}.} underscoring the need for the reassessment of standard techniques intended to examine stellar galactic components (e.g. assumptions on the morphology of the stellar disk, which are typically assumed exponential). One key question yet to be addressed is whether the exponential nature of the disk is conserved within the galactic centre. This conventional assumption implies that the spectro-photometrical properties and dynamics of the disk are conserved, i.e. (a) there is no significant variation of the stellar populations and specific star-formation rate throughout the entire disk and (b) bulge, disk, and bar (if present) co-exist in the galaxy centre without significant dynamical interaction and mass exchange over the course of several gigayears. Both implications, however, appear to be in contrast with indications from recent work, as discussed bellow. 

Through the development of a spectro-photometric bulge-disk decomposition technique and based on its application to a representative sample of 135 local LTGs from the Calar Alto Legacy Integral Field Area (CALIFA) survey \citep{San16}, \cite{Bre20b} suggest that about a third of the spiral disks show a central flattening or even strong down-bending. This result is consistent with previous theoretical and observational studies that further support the existence of a deficit, or even a complete depletion, of the rotation-dominated stellar populations in the centre of a subset of spiral galaxies \citep{Obr13,Du20,Zhu18b}, which might originate at early phases of galaxy formation or reflect the cumulative impact of the dynamically hot bulge on the dynamically colder (and more fragile) disk, promoting angular momentum transfer between the two structural components over several gigayears. A down-bending of the disk within the galactic centre would imply substantial conceptual changes in our understanding of galaxy formation and evolution. For instance, bulge luminosities are expected to increase across the galaxy mass range, with a stronger impact at intermediate-to-low masses \citep{Pap22}. Consequently, central disk depletion may affect the black hole versus bulge mass relation and reveal substantial bulges in systems previously classified as bulgeless. Our attempt to investigate the disk's radial distribution within the bulge in this pilot study further indicates either a pronounced reddening of the inner disk or a significant reduction of rotationally supported stellar populations in galaxy centres.

In addition, recent observational studies have revealed that the nuclear regions of spiral galaxies are far more complex than a simple bulge plus disk (plus bar) model would suggest. In many barred galaxies, nuclear rings -- formed by gas accumulating near resonant radii such as the inner Lindblad resonance -- have been observed to both ignite star formation and channel material towards the super massive black hole \citep[e.g.][]{Agu16,Lin17,Sch23}. Equally intriguing are nuclear disks, which appear to form through prolonged gas inflow and dynamical settling, being often associated with inner bars (although not exclusively) that funnel material inwards in a `galaxy within a galaxy' scenario \citep[e.g.][]{Gad20,Bit21}. 
In this regard, the advent of integral field spectroscopy (IFS) has enabled the systematic identification of composite bulges in an increasing number of disk galaxies, where hot bulges coexist with an embedded cold, rapidly rotating nuclear disk \citep[see e.g.][]{Fal06}. In these systems, the CB dominates the central velocity dispersion, while the nuclear disk is characterised by distinct rotational signatures and an h$_{3}$ velocity anti-correlation \citep[e.g.][]{Fab12,Men14,Erw15,Gad20}. The presence of bars appears to play a significant role in the formation of these composite structures, as bars can drive gas inflows that contribute to the buildup of central disky components (\citealt{Gad25}; see \citealt{Sch25} for a review), even though nuclear disks are often identified in non-barred systems. However, the exact mechanisms determining why some disk galaxies develop composite bulges while others do not remain to be identified, with factors such as the galaxy's merger history, bar presence or strength, and gas content potentially influencing the outcome.

In this work, we apply \textsc{GLANCE} (Galactic archaeoLogy via chronochemicAl \& dyNamiCal modElling), a python framework for automated analysis of IFS data introduced in \cite{gla}\footnote{We refer to this work for a comprehensive description of the adopted methodology.} to a sample of eight disk galaxies with exceptionally high quality IFS data observed with the Multi Unit Spectroscopic Explorer (MUSE). 
Our investigation yields detailed insights into the stellar components of these galaxies, placing particular emphasis on their central regions, where we uncover an intricate landscape. While we acknowledge that the present study is based on a small sample of spiral galaxies hosting intermediate- to massive bulges, these results do hint at formation histories that are more complex than suggested by the dichotomy scenario. In addition, considering that exponential disk extrapolation and subtraction is a widely adopted practice across all late-type systems, our results motivate a reassessment of assumptions on the conservation of the disk’s exponential profile in galaxy centres.

\section{Data sample and methodology\label{meth}}

The sample comprises eight spiral galaxies observed with the MUSE integral field spectrograph drawn from multiple surveys, namely PHANGS, \citep{phangs}, MAD, \citep{mad}, TIMER, \citep{timer}, and AMUSING, \citep{amusing}. Table~\ref{tab1} displays galaxy identifiers, associated MUSE and K-band photometry surveys, logarithmic dynamical masses at the effective radius (log M$_{\rm dyn,e}$/M$_\odot$), target signal-to-noise ratios (tS/Ns) for spatial binning, and environmental data.

\iffalse
\begin{table}[h]
\centering
\begin{tabular}{>{\centering\arraybackslash}m{1.7cm}>{\centering\arraybackslash}m{1.5cm}>{\centering\arraybackslash}m{1.5cm}>{\centering\arraybackslash}m{0.7cm}>{\centering\arraybackslash}m{0.5cm}>{\centering\arraybackslash}m{0.5cm}
>{\centering\arraybackslash}m{0.5cm}}
Galaxy & MUSE Survey & K-Phot. Survey & log M$_{\rm dyn,e}$ & tS/N & Ngal & $\sigma_{\star}$\\
\hline
NGC 3512 & AMUSING & UHS    & 9.64   & 50 & 2 & 81.3\\
NGC 1285 & AMUSING & VHS    & 10.15  & 60 & 1 & 93.4\\
NGC 7364 & AMUSING & UKIDSS & 10.50  & 60 & 1 & 147.1\\
NGC 5248 & TIMER   & VHS    & 10.24  & 100 & 3 & 91.7\\
NGC 3521 & MAD     & VIKING & 10.29  & 260 & 1 & 126.0\\
IC 2560  & MAD     & VHS    & 10.39  & 100 & 5 & 117.1\\
 NGC 1566 & PHANGS  & VHS    & 10.38  & 200 & 13 & 241.9\\
NGC 0863 & AMUSING & UKIDSS & 10.78  & 80 & 2 & 312.5\\
\hline
\end{tabular}
\fi
\begin{table}[h]
\caption{Galaxy sample properties.}
\centering
\setlength{\tabcolsep}{1pt}
\begin{tabular}{ccccccc}
Galaxy & MUSE & K-Phot. & $\log M_{\rm dyn,e}$ & tS/N & Ngal & $\sigma_{\star}$\\
\hline
NGC 3512 & AMUSING & UHS    & 9.64   & 50 & 2 & 81.3\\ 
NGC 1285 & AMUSING & VHS    & 10.15  & 60 & 1 & 93.4\\ 
NGC 5248 & TIMER   & VHS    & 10.24  & 100 & 3 & 91.7\\
NGC 3521 & MAD     & VIKING & 10.29  & 260 & 1 & 126.0\\
IC 2560  & MAD     & VHS    & 10.39  & 100 & 5 & 117.1\\
NGC 1566 & PHANGS  & VHS    & 10.38  & 200 & 13 & 241.9\\
NGC 7364 & AMUSING & UKIDSS & 10.50  & 60 & 1 & 147.1\\
NGC 0863 & AMUSING & UKIDSS & 10.78  & 80 & 2 & 312.5\\
\hline
\end{tabular}
\tablefoot{Identifiers, parent MUSE surveys, K-band photometry sources (VISTA: VHS/VIKING; UKIRT: UKIDSS/UHS), logarithmic dynamical masses at the effective radius obtained with \dyn, target continuum signal-to-noise ratios for Voronoi binning, number of galaxies in the host group (environment information provided by \citealt{Tem16}), and light-weighed average stellar velocity dispersion within the bulge.}
\label{tab1}
\end{table}

To analyse the sample, we used the \textsc{GLANCE} modules for surface photometry, spectral fitting, emission-line analysis, kinematic extraction, and orbit-based dynamical modelling. Assessment and subsequent subtraction of the nebular continuum contribution was carried out prior to spectral synthesis with \starlight\ and kinematics extraction. Photometric decomposition was performed by fitting a Sérsic model to the central excess of the surface brightness profile (SBP) extracted from $K$-band VISTA/UKIRT data while accounting for point spread function (PSF) convolution effects. Considering that at the time of this analysis the adopted \dyn\ implementation was not optimised for strongly barred systems, barred galaxies were visually identified and excluded from the sample, with the exception of IC 2560, as \dyn's best-fitting models agree remarkably well with the data. Spectral analysis of the MUSE datacubes was performed as follows:

%\vspace{-0.5cm}
\begin{itemize}
    \item Voronoi binning \citep{voronoi}, achieving the tS/N values listed in Table~\ref{tab1};
    \item Stellar population synthesis with \fado\ \citep{fado} and \starlight\ \citep{starlight};
    \item Kinematics extraction via \ppxf\ \citep{ppxf};
    \item Dynamical modelling with the \dyn\ orbit-superposition technique \citep[e.g.][]{dyn}, followed by orbital decomposition using cut circularity ($\lambda_{\rm z}$) values of [0.8, 0.25, -0.25] defining the hot, warm, cold, and counter-rotating components.
\end{itemize}

\section{Shedding light on the complex architectures of disk galaxy centres\label{res0}} 

The ultimate goal of this study is to uncover the true radial luminosity profile of the disk in a subset of spiral galaxies, ideally obtaining a more realistic functional form of the disk as compared to the overly simplified exponential profile typically adopted. However, the small sample size prevented us from drawing definitive conclusions, as the exceptional data quality posed significant challenges for dynamical modelling, often impairing convergence to robust solutions. Nevertheless, by employing \dyn\ to model the observed kinematics and decompose stellar orbits into cold, warm, and hot components, the analysis uncovered a striking degree of structural complexity in the central regions of these systems (we refer to the appendix for detailed results of the analysis for each galaxy of the sample). Interestingly, with the exception of one galaxy in our sample (whose cold component exhibits an approximate but imperfect match to the typically assumed exponential profile), the systems exhibit one of two distinct behaviours in their cold component: a sharp central drop forming a doughnut-shaped structure, which might be primordial or result from posterior orbital heating, or the presence of a compact inner disk that is significantly steeper than the outer main disk embedded within the kinematically hot component.

\begin{figure*}[h!]
\centering
\includegraphics[width=\linewidth]{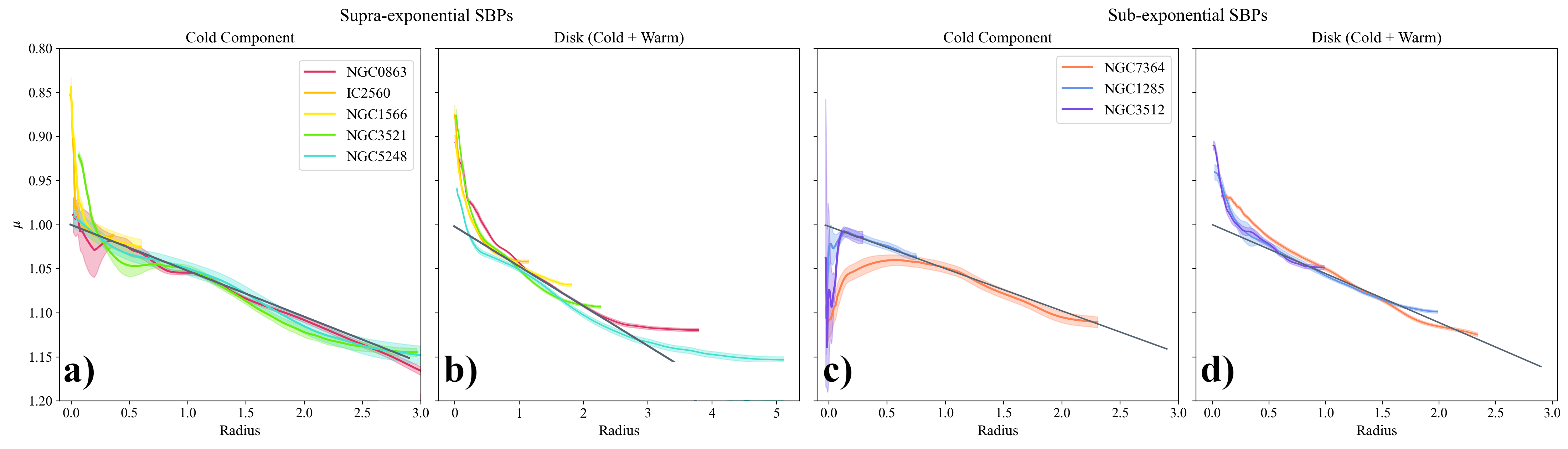}
\caption{Surface brightness profiles of the cold (panels a and c) and cold plus warm (panels b and d) orbital components for the galaxy sample normalised by each galaxy’s best-fit exponential disk's scale length and central surface brightness. Shaded regions indicate 1$\sigma$ uncertainties, derived from models whose $\chi^2$ values lie within 1$\sigma$ of the best-fit solution. Each line is coloured according to total galaxy mass (reddest curves indicate the most massive systems, bluest the least). Two rightmost panels (`supra-exponential SBPs'): Galaxies whose cold profiles rise above the exponential in their innermost regions, revealing a `nuclear disk'' within $\sim$0.5 R$_{\rm D}$. Two leftmost panels (`sub-exponential SBPs'): Galaxies whose cold profiles descend below the exponential (doughnut morphology) within the same region. The standard exponential profile is shown in black. Different colours represent the various galaxies of the sample.}
\label{Fig3}
\end{figure*}

The morphology of the central cold component can be appreciated from Fig.~\ref{Fig3}, which displays the normalised cold-orbit and disk (cold plus warm) SBPs\footnote{These SBPs are derived from orbit-based mass fractions converted into light using the best-fitting mass-to-light ratio and the total MUSE luminosity.} for the eight sampled galaxies, with radius expressed in units of each galaxy's exponential scale length (R$_{\rm D}$) and surface brightness $\mu$ normalised to the central extrapolated value. Solid curves trace the radial distribution for each galaxy, with respective shaded bands indicating 1$\sigma$ uncertainties as derived from models whose $\chi^2$ values lie within 1$\sigma$ of the best-fit solution. To facilitate direct comparison, the sample is divided according to the inner morphology of the cold component: Galaxies exhibiting a supra-exponential central upturn, being associated with the presence of a nuclear disk embedded within the bulge are displayed in the two rightmost panels, and those showing a sub-exponential central decline, indicating a relative depletion of cold orbits in the same region, are in the two leftmost panels. NGC 5248, whose cold component shows a rough resemblance to the pure exponential profile, is included in the first group. Note that while panels a and c show only the cold components, panels b and d present the combined cold plus warm (often assumed to correspond to the total disk) profiles. Interestingly, inspection of panels b and d reveals that the total disk components of all galaxies in the sample significantly deviate from a simple exponential, resembling instead a S\'ersic profile with index $\eta$ > 1\footnote{We refer to the online appendix at \href{https://doi.org/10.5281/zenodo.19356335}{Zenodo} for the Sérsic fits to the disk component.}, in agreement with the profiles found in TNG50-simulated disk galaxies \citep{Du21}.

\begin{figure*}[!b]
\centering
\includegraphics[width=1\linewidth]{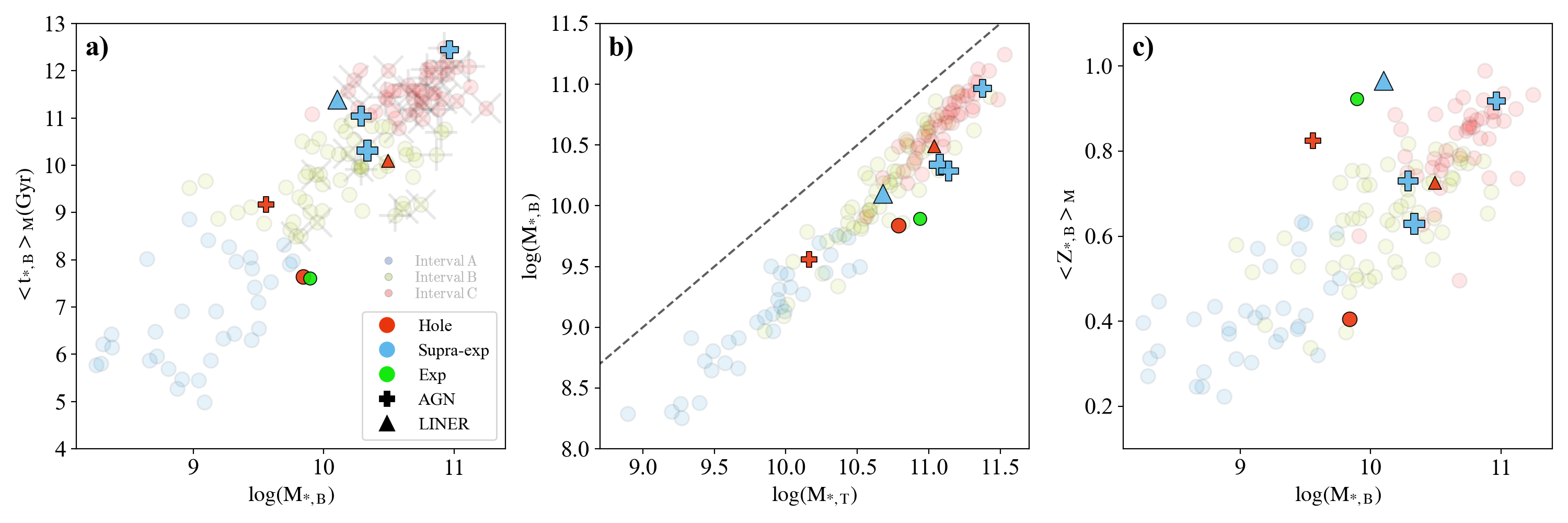}
\caption{Comparison of the bulge properties between the reference sample and the galaxies analysed in this work (coloured symbols). %The reference sample consists on 135 face-on disk galaxies from the CALIFA survey, being marked by transparent circles (featuring interval A, B, and C galaxies), where bulges are subdivided into the three intervals according to their $\langle\delta\mu_{9\mathrm{G}}\rangle$ values, which reflect the fractional contribution of stellar populations younger than 9~Gyr, the highest of which is in iA bulges while being intermediate in iB and lowest in the oldest, most evolved iC systems)
The reference sample consists of 135 face-on disk galaxies from the CALIFA survey, shown as transparent circles (including iA, iB, and iC systems): Bulges are classified into these three intervals according to their $\langle\delta\mu_{9\mathrm{G}}\rangle$ values, which trace the fractional contribution of stellar populations younger than 9~Gyr, being highest in iA, intermediate in iB, and lowest in the most evolved iC systems.
Panel (a): Relationship between bulge stellar mass (\mbstar) and mass-weighted stellar age (\mtmass). Panel (b): Total stellar mass (\mstotal) versus bulge stellar mass. Panel (c): Bulge stellar mass versus bulge mass-weighted metallicity (\mZmass), derived using \starlight, adopting the Z4 library. The vivid markers are colour coded according to the structure of their central cold component (with red, green, blue indicating centrally depleted, exponential, and supra-exponential, respectively). The shape of the markers represents nuclear activity (i.e. circles correspond to non-active galaxies, crosses to AGNs, and triangles to LINERS), and the size indicates the local environmental density.}
\label{Fig1}
\end{figure*}

Given the limited size of our sample, we contextualised our results by comparing them to a reference set of 135 disk galaxies from the CALIFA survey \citep[][BP18]{BP18}, which are representative of the broader population of spiral galaxies. In that study, different techniques such as optical surface photometry, full spectral modelling of CALIFA IFS data, and \RY, a post-processing tool to remove the light contribution of young stellar populations \citep{RY}, were combined, yielding scaling relations between bulge and disk physical properties, the total galaxy mass, and the bulge-to-total ratio. In the present work, we extended beyond spectro-photometric techniques by including spatially resolved kinematics and Schwarzschild dynamical modelling, offering a comprehensive view of the galaxies’ inner stellar structures. Apart from deriving the mean stellar age and metallicity of the bulge and disk, we also mapped their orbital structures into cold, warm, and hot. We note that beyond the kinematic and dynamical analysis introduced here, the main methodological differences relative to BP18 can be summarised as follows:

%\vspace{-0.5cm}
\begin{itemize}
    \item Photometric band and resolution: The CALIFA bulge radii were derived from SDSS r-band imaging, which systematically yields larger values compared to our K-band-based measurements. This offset likely arises from PSF effects, where the sharper near-infrared PSF of K-band data resolves more compact structures, and the K-band’s sensitivity to older stellar populations, which tend to be more centrally concentrated.
    \item Spectral coverage: The MUSE spectral range starts at $\sim4750\,\AA$, while CALIFA provides coverage down to 4000 \AA, partially including the UV regime. To quantify the resulting bias, we re-ran the \textsc{STARLIGHT} spectral synthesis code on the CALIFA bulge integrated spectra while excluding this wavelength range. Though mass-weighted stellar ages remain consistent between the two configurations ($\Delta t_{\rm M} < 0.05$\,Gyr), omitting this wavelength interval tends to systematically increase assessed mass-weighted stellar metallicities (see Fig.~\ref{Z_BP18}), seemingly consistent with the disagreement shown in Fig.~\ref{Fig1}c.
\end{itemize}

\begin{figure}[!t]
\centering
\includegraphics[width=1\linewidth]{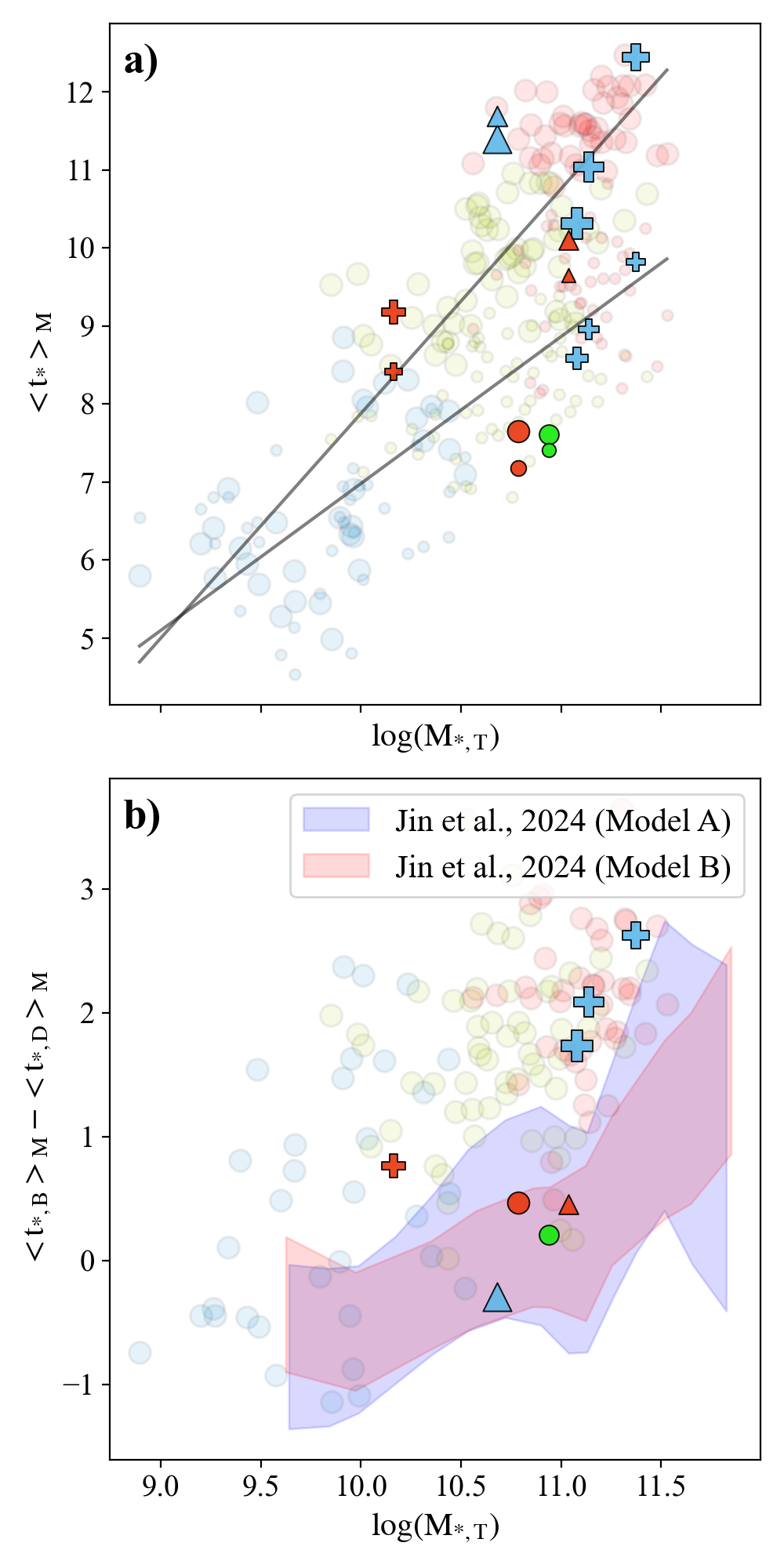}
\caption{Panel (a): Relationship between \mstotal\ and \mtmass\ (large markers) and the parent disk (\mdtmass, small markers). Panel (b): Comparison of \mstotal\ and the mean age difference between bulge and disk components. In panel (b), the regions shaded blue and red depict the age difference between the hot and cold orbit components within the effective radius as a function of stellar mass for two different models, as derived by \citet{Jin24}. Markers follow the same coding as in Fig.~\ref{Fig1}.} 
\label{Fig2}
\end{figure}

Figures \ref{Fig1} and \ref{Fig2} compare the stellar population properties of bulges and disks in our sample against the CALIFA reference sample. Figure \ref{Fig1} illustrates the relationship between the present-day bulge stellar mass \mbstar\ and (panel a) the mass-weighted stellar age \mtmass, (panel b) the total stellar mass \mstotal, and (panel c) the mass-weighted stellar metallicity \mZmass. As anticipated from the methodological differences previously outlined, some galaxies in our sample exhibit a higher \mZmass\ compared to CALIFA (see Fig.~\ref{Fig1}c), 	probably attributable to the total exclusion of the UV wavelength regime in the spectral fitting of the MUSE sample. Other parameters, including \mbstar\ and \mtmass, remain consistent between the two samples. Figure~\ref{Fig2} further contrasts the mass-weighted stellar ages of bulges and their host disks, with panel (b) additionally displaying shaded blue and red regions extracted from  Fig. 11 of \citet{Jin24}. These estimates were obtained by tagging the stellar ages of 82 CALIFA disk galaxies onto their Schwarzschild-modelled orbits to recover age differences between the cold (disk-like) and hot (bulge-like) components. Although this approach is fundamentally different from the one taken by BP18, both studies agree that the bulge-disk age offset is negative at low total stellar masses and increases with mass, with the models from \citet{Jin24} yielding lower age differences for higher mass galaxies, as compared to the results by BP18. For the present sample, marker colours encode the morphology of the cold stellar component: blue for nuclear disks, green for structures resembling an exponential disk throughout the full galaxy's extent, and red for central depletion (doughnut-shaped cold component). Furthermore, crosses and triangles flag active galactic nucleus (AGN) and low-ionization nuclear emission-line region (LINER) hosts, respectively, with the marker size reflecting the local galaxy density. The reference sample is shown as transparent points, colour-coded by the fractional contribution of old ($\gtrsim$ 9 Gyr) stellar populations within the bulge (blue: negligible, green: intermediate, red: dominant). The AGN and LINER hosts are marked with X’s and crosses, respectively. 
Additional information is provided in the appendix, including results from surface photometry (\ref{photSBP_3}), spectral synthesis with FADO (\ref{fado0_3}), dynamical modelling and associated diagnostic plots (\ref{dyn2_1} and \ref{dyn2_2}), and SBPs derived from the dynamical decomposition (\ref{dynSBPs_3}). The remaining galaxies are presented in the online appendix.

The trends in both figures align with established correlations: More massive galaxies host bulges that are older, more metal-rich, and proportionally more massive relative to their disks. However, no clear connection emerged between these trends and the morphology of the cold component. Intriguingly, galaxies exhibiting a depleted central cold component tend to be isolated or reside in small groups, whereas composite bulges -- those hosting a nuclear disk embedded in a kinematically hot structure -- are preferentially older and reside in denser environments,  with NGC~3521 being the only isolated exception. 

\section{Discussion and conclusions\label{con}}

Although the limited size of our sample prevents us from drawing definitive conclusions, the exceptionally high data quality has uncovered a complex and diverse landscape of stellar structures in the centres of spiral galaxies. This challenges two prevailing premises: first, that the formation of disk galaxies follow one of two distinct pathways and, second, the common assumption that the stellar disk can always be adequately described by a simple exponential profile extending into the bulge-dominated region.

Regarding the first premise, the bimodal CB versus PB scenario appears overly simplistic in light of the morphologies of central structures observed in our sample. 
The majority of our oldest, most massive bulges, while exhibiting high $\sigma_{\star}$ indicative of pressure support, simultaneously host compact nuclear disks. Several studies have provided evidence of such structures through photometry, kinematic modelling, and stellar population analyses \citep[e.g.][]{Fab12,Men14,Erw15,Gad20,Gad25} not only at centres of spirals but also embedded within dwarf ellipticals \citep[e.g.][]{Jan12,Jan14,Tol15}. Interestingly, our own Milky Way, which is not expected to be atypical among spiral galaxies but uniquely well resolved, also displays a remarkable degree of central complexity, featuring a nuclear disk with diverse stellar populations coexisting with bulge and bar populations \citep[e.g.][]{Nog22,Nog23,Schu26}. Concerning the observed depletion of cold orbits in the central regions of some galaxies in our sample, \cite{Zhu18b} has reported similar central deficits in the dynamically cold component in previous Schwarzschild-based orbital decompositions of nearby galaxies.

\begin{figure*}[!t]
\centering
\begin{minipage}{0.65\linewidth}
    \includegraphics[width=\linewidth]{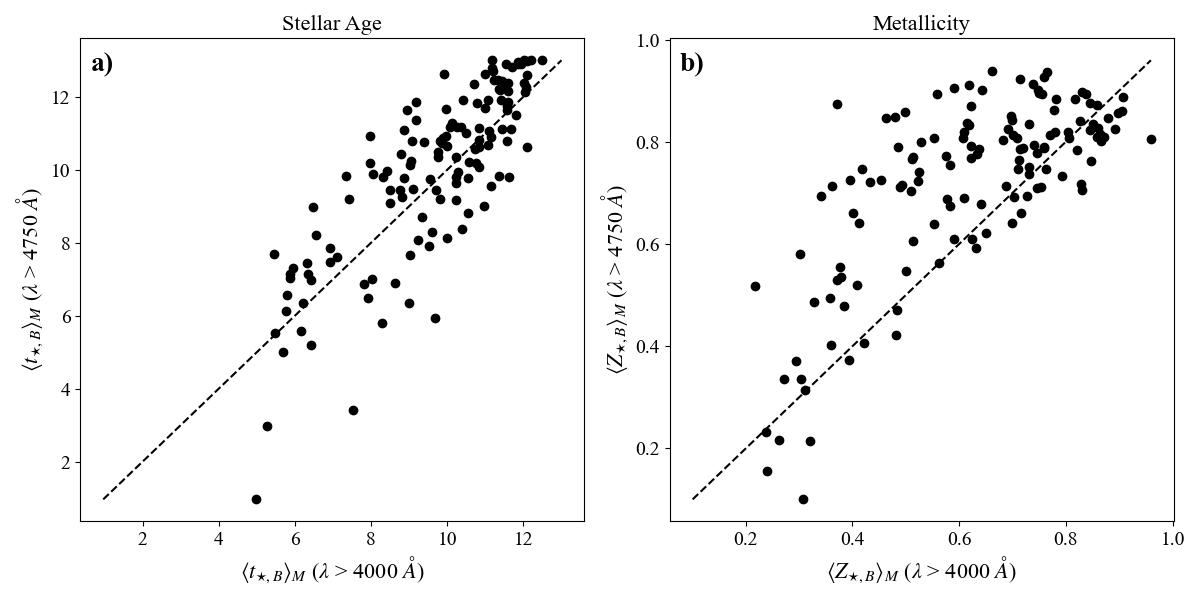}
\end{minipage}
\hfill
\begin{minipage}{0.3\linewidth}
    \caption{Comparison between \mtmass\ (panel a) and \mZmass\ (panel b) for the bulges of the reference sample (135 CALIFA disk galaxies) derived using \starlight\ (Z4 library). The x-axis shows results obtained when partially including the UV regime (4000 - 4750\AA), while the y-axis displays the results when excluding it (both have spectral coverage extending to 6800\AA). While stellar age estimates remain consistent, stellar metallicity tends to increase when this wavelength range is omitted. This systematic shift may explain why some metallicity assessments from MUSE data (which totally lack the UV regime) yield higher values compared to the reference sample.}
    \label{Z_BP18}
\end{minipage}
\end{figure*}

It is nevertheless important to acknowledge that orbit-superposition inferences can be sensitive to modelling choices and methodological implementation. Mock-galaxy recovery tests with \textsc{DYNAMITE} performed in \cite{San24} indicate median systematic offsets of the order of 0.05--0.15 in the recovered cold, warm, hot, and counter-rotating orbital fractions within $\sim$1R$_{\rm e}$, with scatter reaching $\sim$0.2--0.25 in individual systems. In addition, variations in the adopted kinematic extraction method can shift global orbital fractions by $\sim$0.1--0.2 in specific cases \citep{Rei25}, while tests of alternative stellar mass-to-light ratio prescriptions indicate that the overall circularity structure is largely preserved \citep{Tha23}. Lastly, tests of Schwarzschild orbit-superposition models using mock observations have shown that the main orbital structures are robustly recovered across different viewing angles, implying that moderate inclination uncertainties mainly lead to small shifts in the relative fractions of cold, warm, and hot orbits without significantly altering their radial distributions \citep[e.g.][]{Zhu18b}. Taken together, these results imply an overall systematic uncertainty of the order of $\sim$10--20\% on individual cold, warm, hot, and counter-rotating fractions, with larger uncertainties expected in the innermost regions. However, these systematic effects are not expected to be sufficiently strong to artificially suppress the cold component to negligible levels nor to generate spurious nuclear disk-like structures in the absence of an underlying dynamically cold population.

The presence of nuclear disks embedded within kinematically hot bona-fide CBs presents an intriguing puzzle, particularly given the overall undisturbed morphologies of the galaxies, as it is unclear why some disk galaxies develop a cold central nuclear component alongside a hot bulge while others do not. 
This apparent contradiction may be tentatively interpreted in terms of two plausible, non-exclusive formation scenarios. First, secular evolutionary processes could have funnelled gas inwards over extended timescales, gradually forming a compact rotating disk within the pre-existing pressure-supported bulge. Alternatively, these systems may have experienced minor mergers or interactions several gigayears ago, which deposited cold gas into the central regions before the galaxy relaxed to its current configuration. In this case, the nuclear disk could represent a relic of past accretion events whose tidal signatures have since faded below detectable levels. Both mechanisms could operate concurrently, with their relative importance potentially depending on the host galaxy's mass and environmental context. Conversely, the majority of our bulges with central stellar depletion in their cold component exhibit a lower $\sigma_{\star}$ and younger mass-weighted ages, having no considerable degree of rotation.

With respect to the second premise, the observed diversity in cold component morphology challenges a fundamental assumption in photometric decomposition, namely that the outer disk structure can be reliably extrapolated inwards to isolate the bulge. For bulges with depleted cold components, the absence of rotating orbits in the galaxy's centre limits the validity of this approach. Likewise, nuclear stellar disks seem to be representing a distinct kinematic entity rather than an extension of the outer disk.
If we instead consider the morphology of the cold plus warm orbital structures, our results reveal a broad range of radial profiles, with S\'ersic indices consistently > 1 (see the SBPs derived from dynamical modelling in the online appendix\footnote{Considering that the cold plus warm SBPs do not always resemble a S\'ersic profile, we obtained reliable fits for only five of our eight galaxies.}). Moreover, even beyond the bulge radius, this combined profile lies systematically below the galaxy’s overall SBP, suggesting that the parent disk -- beyond the central region of the galaxy -- hosts a non-negligible fraction of hot and/or counter-rotating orbits. 

Taken together, these and similar findings from equally excellent quality data and more sophisticated contemporaneous analysis techniques suggest that both the common practice of inwardly extrapolating the outer disk and the conventional bimodal scenario for the formation and evolution of disk galaxies may be overly simplistic or incomplete. Speculatively, one way to reconcile these results could be to consider that the cumulative impact of low-mass mergers over a disk galaxy’s lifetime may play a key role in both shaping its central structure and blurring the distinction between classical and PBs into a continuous sequence with mass. In this scenario, the number and angular momentum of such accretion events, relative to those of the host galaxy, may determine whether the accumulated material settles into nuclear disks, which in our admittedly small sample are mostly found within massive, dynamically hot bulges hosted by more massive galaxies and typically located in denser environments. The composite structure of the parent disk, which seems to contain a noticeable fraction of hot or counter-rotating orbits, could also naturally result from cumulative minor-merger heating.

Obtaining a comprehensive understanding of the formation and evolution of LTGs (and determining whether this or other proposed scenarios drive the observed structural complexity) can be achieved through high signal-to-noise spatially resolved spectroscopy of a statistically representative sample of local disks. These data will allow for robust characterisation of the kinematic and stellar population properties of the different stellar components, thus paving the way for decomposition models that move beyond the simplistic exponential-disk assumption. Concurrently, the exploration of high-resolution cosmological simulations such as TNG50 may provide valuable insights into these processes by tracing the interplay between mergers, gas inflows, and internal dynamical evolution that leads to the emergence of composite bulges.

\section*{Data availability}

The online appendix, available at https://doi.org/10.5281/zenodo.19356335, presents the full set of results for the remaining galaxies, including surface photometry, spectral synthesis (FADO), dynamical modelling outputs with diagnostic plots, and the corresponding surface brightness profiles from the dynamical decomposition.

\begin{acknowledgements}
I.B. has received funding from the European Union's Horizon 2020 research and innovation programme under the Marie Sklodowska-Curie Grant agreement ID n.º 101059532. This project was extended for 6 months by the Franziska Seidl Funding Program of the University of Vienna.
J.F-B acknowledges support from the PID2022-140869NB-I00 grant from the Spanish Ministry of Science and Innovation.
M. O. is supported by JSPS KAKENHI Grant Number JP25K07361. 
I.P. acknowledges funding by the European Research Council through ERC-AdG SPECMAP-CGM, GA 101020943.
We would also like to thank Dimitri A. Gadotti for his valuable comments and suggestions. 
This research has made use of the NASA/IPAC Extragalactic Database (NED) which is operated by the Jet Propulsion Laboratory, California Institute of Technology, under contract with the National Aeronautics and Space Administration.
\end{acknowledgements}

% References

% Appendix
%\newpage
\appendix
\onecolumn

%\newcommand{\plainapphead}[1]{%
%  \par\leavevmode\par
%  \noindent{\large\bfseries #1}\par}

%\pagestyle{plain}

%\onecolumn

%\appendix
%\renewcommand\chaptername{Appendix}

\section{Supplementary maps}\label{app}
%\vspace{0.5cm}

% -------------------------------------------------

This appendix provides a brief description of the galaxies, the assessment of the bulge radius via surface photometry, a subset of maps derived by \textsc{FADO}, and diagnostic material enabling a visual assessment of the \textsc{DYNAMITE} fits, respectively. Corresponding material for the remaining systems can be found in the online at \href{https://doi.org/10.5281/zenodo.19356335}{Zenodo}, with exception of NGC~1566 which results are presented in \citealt{gla}.

$\bullet$ ~\textbf{NGC 3521}: Flocculent intermediate spiral galaxy with an inclination of $\sim$70$^\circ$, being the most inclined galaxy of our sample. Dynamical analyses reveal a prominent nuclear disk embedded within its bulge (already evident in the kinematic maps), while FADO-derived maps show a compact central structure with enhanced luminosity-weighted stellar age, together with a more extended pattern of higher luminosity-weighted stellar metallicity, and lower EW(H$\alpha$) and EW(H$\beta$). It hosts a LINER nucleus, suggesting low-ionisation activity. NGC 3521 is isolated.

\vspace*{-0.3cm}
\subsection{Surface photometry}

\refstepcounter{figure}\label{photSBP_3}
\centerline{\includegraphics[trim={2cm 0 1.5cm 0}, clip, width=0.7\linewidth]{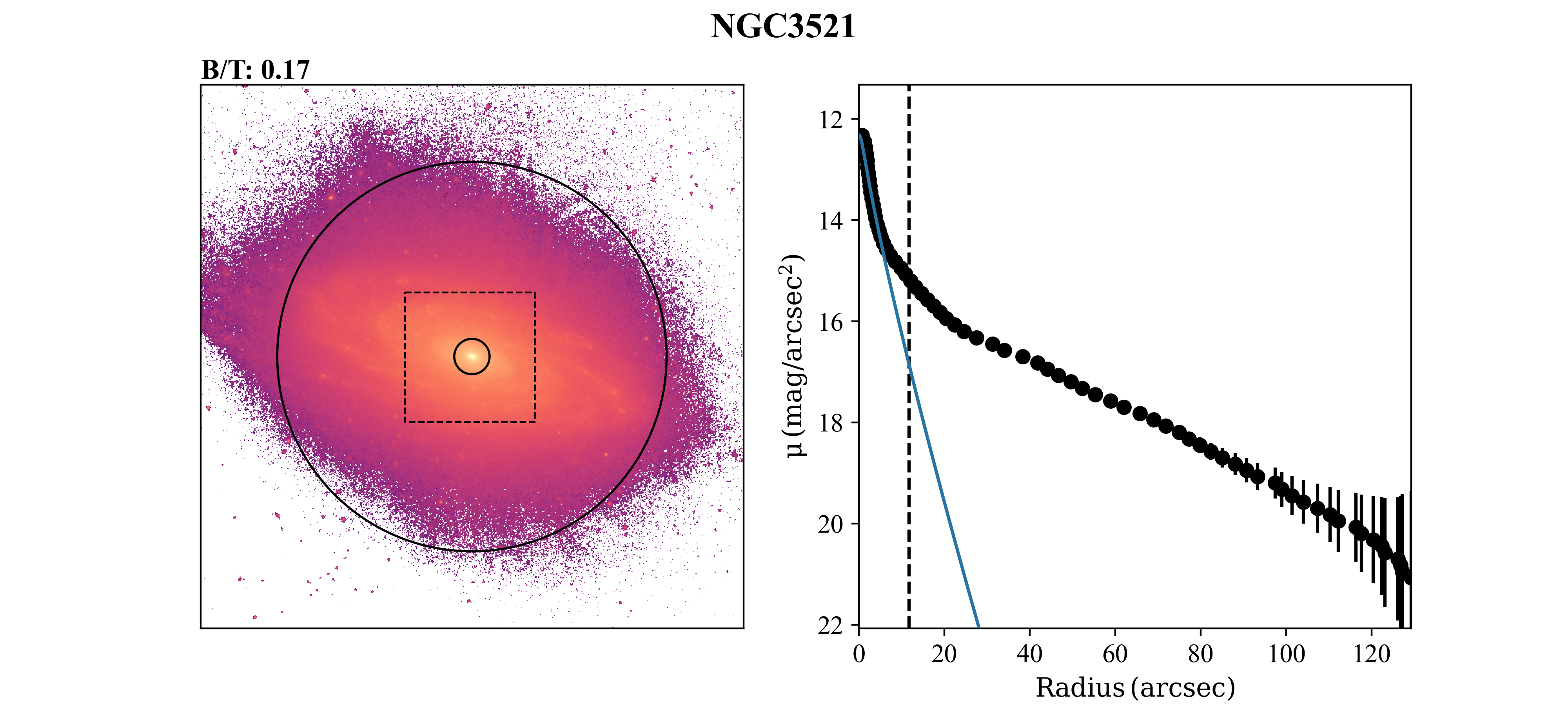}}

\noindent\textbf{Fig.~\thefigure.} Left panel: K-band photometric image of the NGC 3521, as observed by VIKING. The small circle indicates the bulge radius as assessed by the S\'ersic fit, the ellipse illustrates the widest point in the SBP, roughly outlining the galaxy's size, and the square represents MUSE's field of view (FoV). Right panel: Derived SBP with a blue line indicating the best-fitting S\'ersic model to the central luminosity excess. The vertical dashed line represents the bulge radius as determined by the S\'ersic fit. Derived bulge-to-total is displayed above the left panel.
\normalsize

\vspace*{-0.3cm}
\subsection{Spectral synthesis}
\vspace*{0.4cm}

\refstepcounter{figure}\label{fado0_3}
\centerline{\includegraphics[width=1\linewidth]{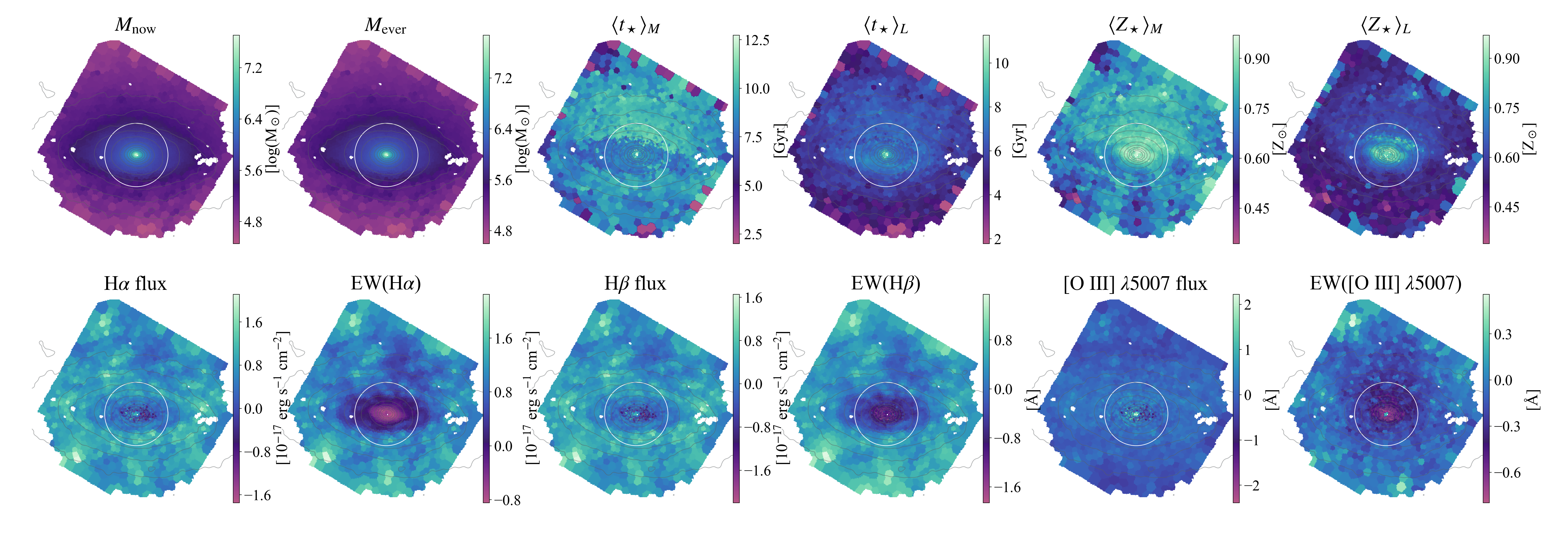}}

\noindent\textbf{Fig.~\thefigure.} Subset of spatially resolved stellar population and nebular emission properties of NGC 3521, derived with \textsc{FADO}, adopting the Z4 SSP Library. Top row (left to right): Present-day and ever-formed stellar mass ($M_{\rm now}$, $M_{\rm ever}$), mass- and light-weighted stellar ages ($\langle t_\star\rangle_M$, $\langle t_\star\rangle_L$), and mass- and light-weighted stellar metallicities ($\langle Z_\star\rangle_M$, $\langle Z_\star\rangle_L$). Bottom row (left to right): H$\alpha$ flux, EW(H$\alpha$), H$\beta$ flux, EW(H$\beta$), [O\,\textsc{iii}]~$\lambda5007$ flux, and EW([O\,\textsc{iii}]~$\lambda5007$). White contours trace the K-band morphology, and the circle indicates the bulge radius.
\normalsize

\vspace*{-2cm}
\subsection{Dynamical modelling}
\vspace*{0.4cm}

\begingroup
\setlength{\parindent}{0pt}
\setlength{\parskip}{0pt}

\refstepcounter{figure}\label{Dyn_maps}
\centerline{\includegraphics[trim={2cm 0 1.5cm 0}, clip, width=0.80\linewidth]{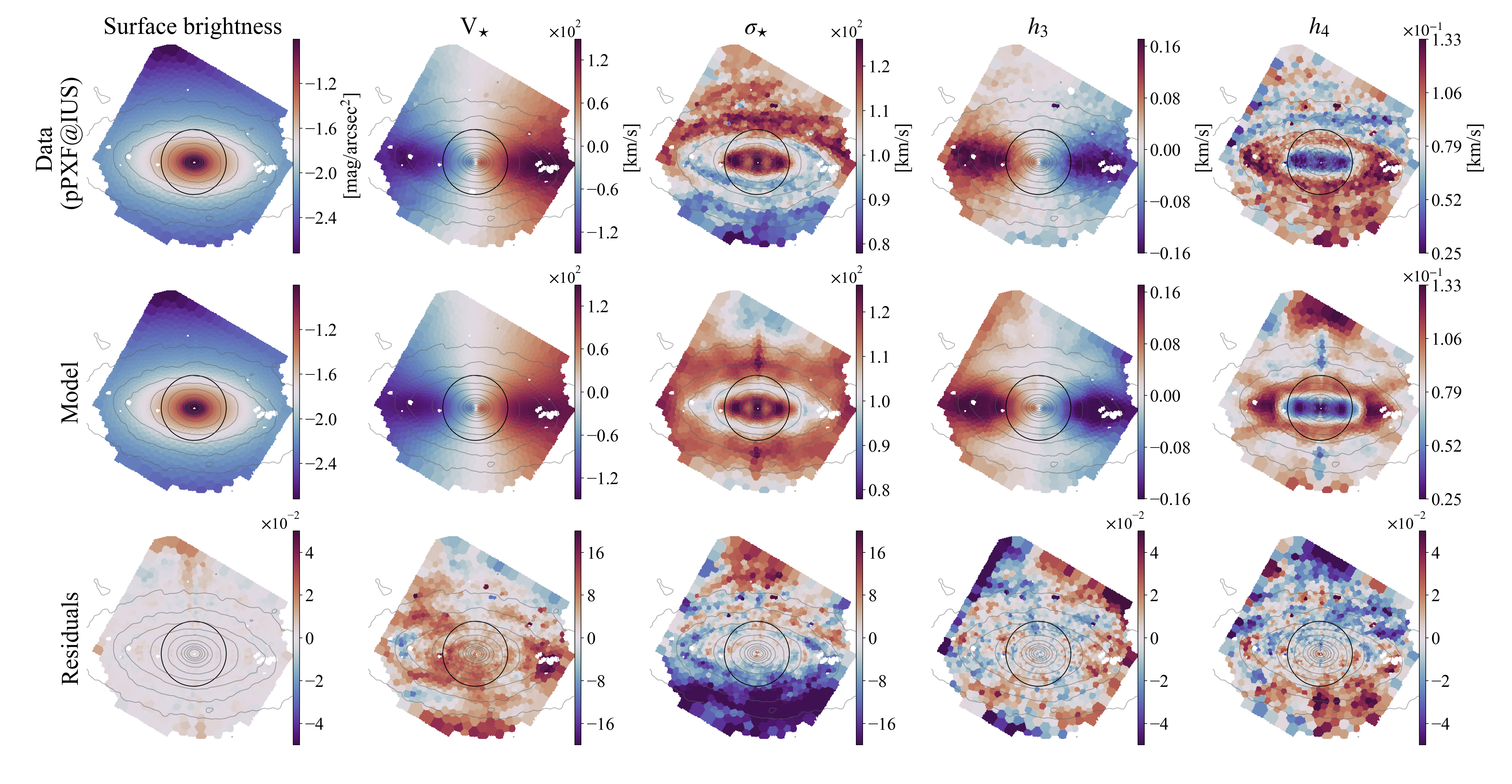}}

\noindent\textbf{Fig.~\thefigure.} Comparison of observed velocity moments (top row), best-fit model moments (middle row), and residuals (bottom row) from the \dyn\ dynamical modelling for NGC 3521. Over-plotted are the K-band contours, as well as a circle defining the bulge radius.
\normalsize

\vspace{6pt}

\refstepcounter{figure}\label{dyn2_1}
\noindent
\begin{minipage}{\linewidth}
\centerline{%
\begin{minipage}[c]{0.48\linewidth}
  \centering
  \includegraphics[width=\linewidth]{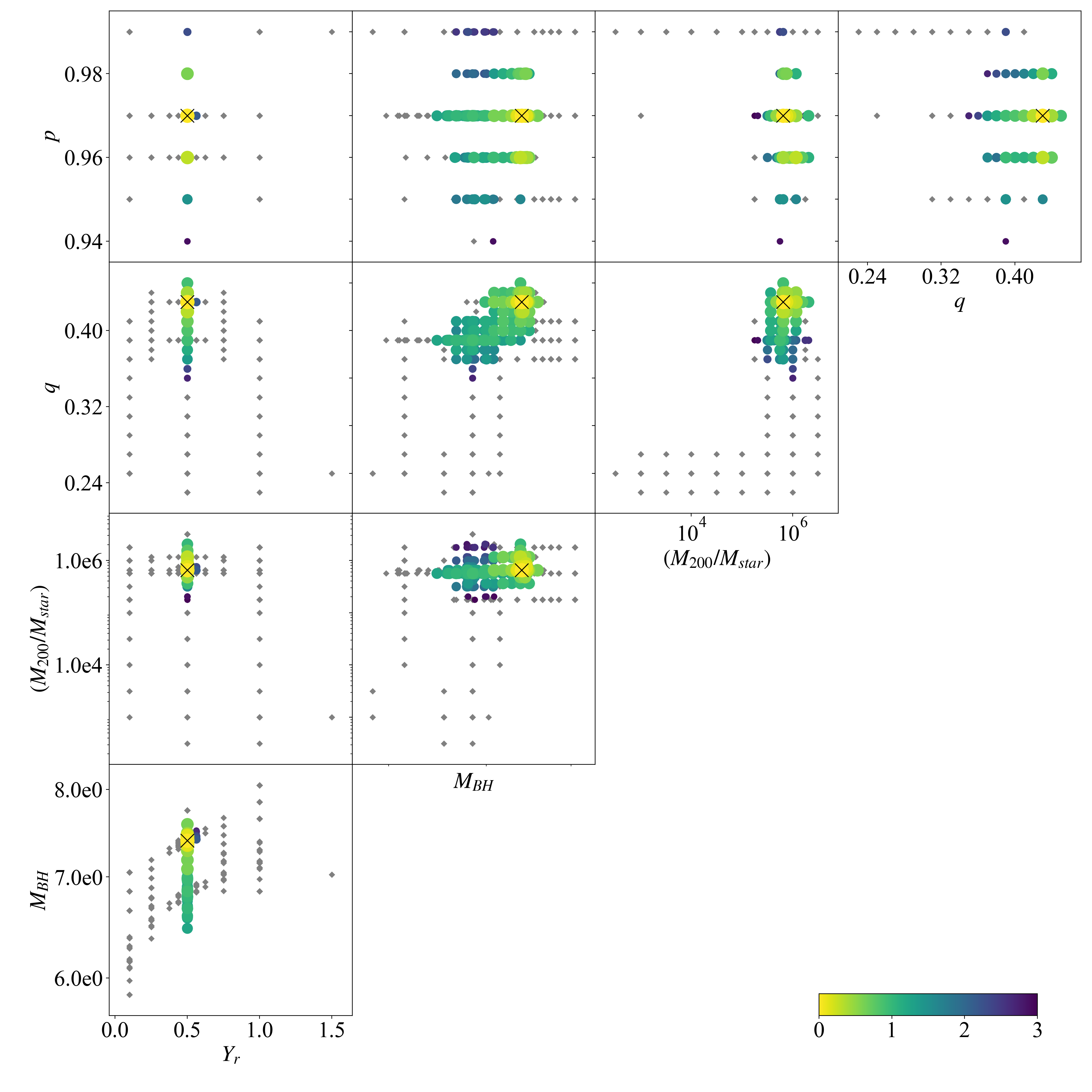}
\end{minipage}\hfill
\begin{minipage}[c]{0.48\linewidth}
  \small
  Relationships between the open parameters ($p$, $q$, $\Upsilon_{r}$,  M$_{200}$, and M$_{\bullet}$) explored during the dynamical modelling process of NGC 3521, with best-fitting values of $p$ = 0.97; $q$ = 0.43; $\Upsilon_{r}$ = 0.5; log(M$_{200}$) = 5.8, log(M$_{\bullet}$) = 7.4.
\end{minipage}}
\end{minipage}
\par

\noindent\textbf{Fig.~\thefigure.} Parameter-space exploration for the dynamical modelling of NGC 3521.
\normalsize

\vspace{6pt}

\refstepcounter{figure}\label{dyn2_2}
\centerline{\includegraphics[width=0.74\linewidth]{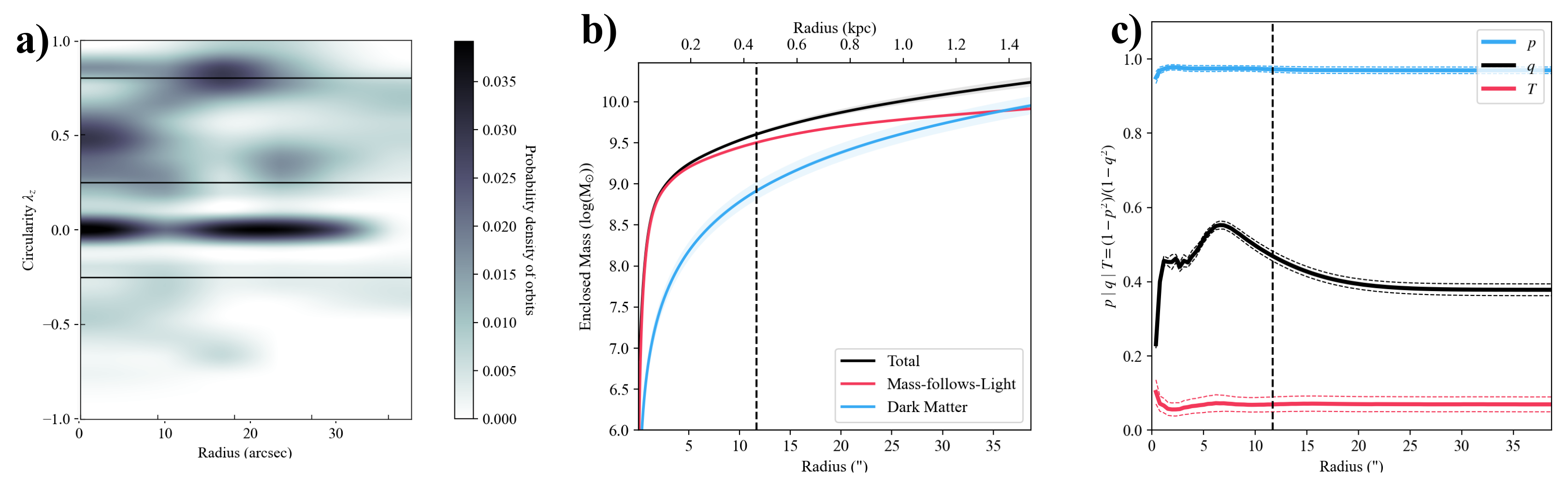}}

\noindent\textbf{Fig.~\thefigure.} Panel (a): Circularity plot showing the distribution of stellar orbits in the best-fit dynamical model of NGC 3521. Horizontal lines denote the separations between cold, warm, hot, and counter-rotating orbits, with default cut values of $\lambda_{z} < 0.8$; $0.8 > \lambda_{z} > 0.25$. Panel (b): Enclosed mass profiles for the stellar (mass-follows-light) component, dark matter, and their sum. Panel (c): Intrinsic axis ratios $q$ and $p$ as functions of distance from the centre; T quantifies triaxiality. Vertical dashed lines indicate the bulge radius.
\normalsize
\endgroup

%\vspace*{-2cm}
\subsection{Orbital decomposition}
\vspace*{0.5cm}

\begin{figure}[!htbp]
\centering
\includegraphics[trim={0cm 0 0cm 1.5cm}, clip, width=0.7\linewidth]{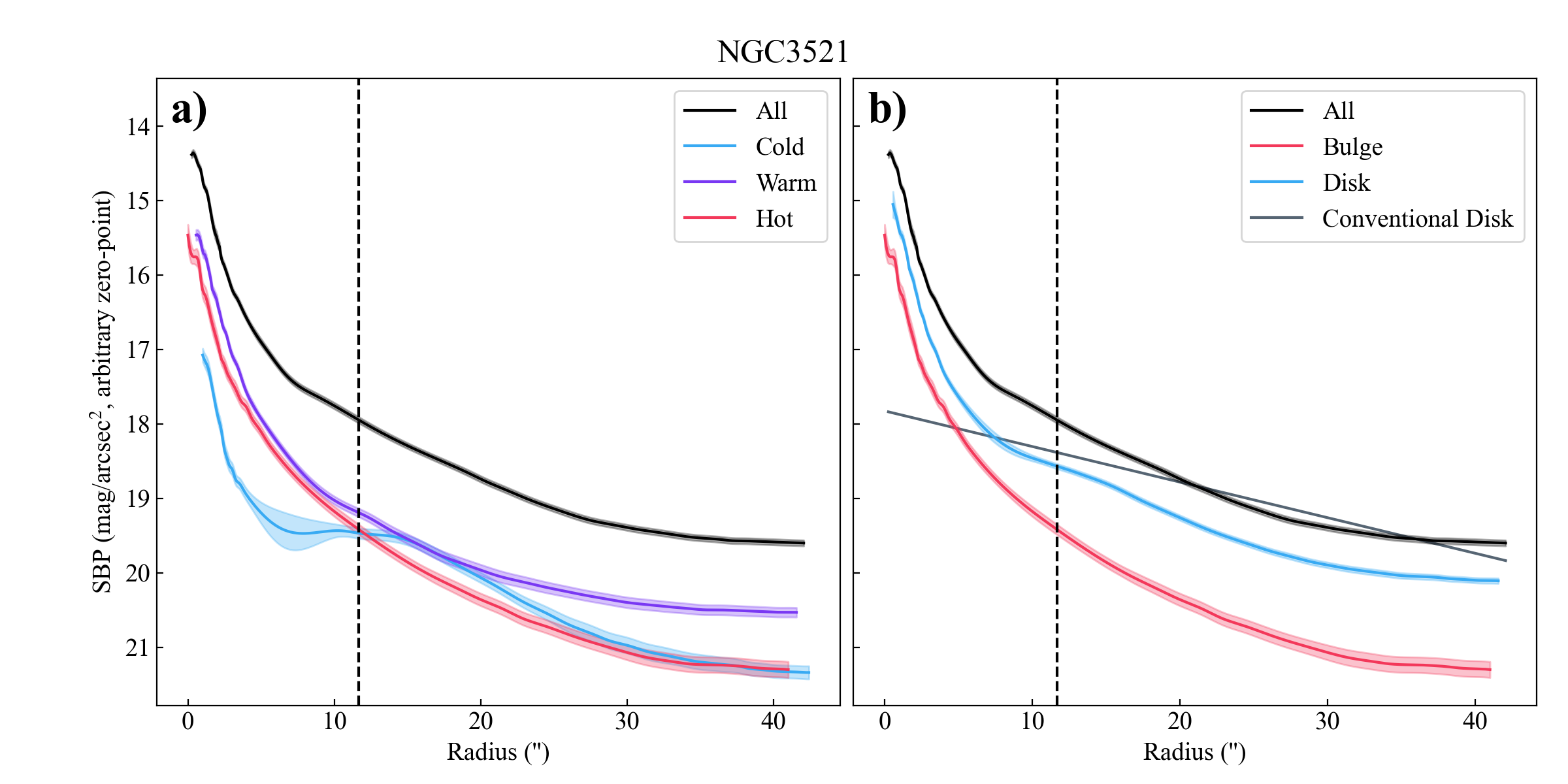}
\caption{SBPs of the different dynamical components of NGC 3521, obtained through dynamical modelling and subsequent decomposition. Shaded regions indicate 1$\sigma$ uncertainties, derived from models whose $\chi^2$ values lie within 1$\sigma$ of the best-fit solution. Panel (a): SBPs of the cold (blue), warm (purple) and hot (red) components. Panel (b): SBPs of disk and bulge, where the disk's is obtained by accounting for both cold and warm orbits, while the bulge corresponds to the hot stellar component. The vertical line depicts the bulge radius as obtained trough K-band surface photometry. Attempts to fit a Sérsic profile to the disk were unsuccessful.}
\label{dynSBPs_3}
\end{figure}

\end{document}